\documentclass[a4paper]{article}
\usepackage{RR}
\usepackage{url} 
\usepackage{hyperref}

\usepackage{amssymb,amsmath}
\usepackage[noend]{algorithmic}
    \renewcommand{\algorithmiccomment}[1]{ \sffamily{//~#1}}
    \newcommand{\algorithmicdeclare}{\textbf{Declare:}}
  \newcommand{\DECLARE}{\item[\algorithmicdeclare]}

\usepackage{algorithm}\algsetup{indent=2em}

\usepackage[OT1]{fontenc}
\usepackage[latin1]{inputenc}

\usepackage{eurosym}
\usepackage{graphicx}
\usepackage{pslatex}
\usepackage{setspace}
\usepackage{stmaryrd}
\usepackage{xspace}

\usepackage{macros}

\RRNo{6069}
\RRdate{Decembre 2006}
\RRauthor{
  Pierre Sutra\thanks{LIP6, 104, ave.\ du Président Kennedy, 75016 Paris, France; \url{mailto:pierre.sutra@lip6.fr}} ~~~ Marc Shapiro\\
  {\small Université Paris VI and INRIA Rocquencourt, France} \\
  \and
  Jo\~{a}o Barreto \\
  {\small INESC-ID and Instituto Superior Técnico, Lisbon, Portugal} \\
}
\authorhead{Sutra \& Barreto  \& Shapiro}
\RRtitle{
  Un protocole de validation pour la réplication optimiste
  dans les systèmes répartis sémantiquement riches
  }
\RRetitle{An Asynchronous, Decentralised Commitment for Optimistic Semantic Replication}
\titlehead{}
\RRresume{%%
  Nous examinons à travers ce document la 
  cohérence dans les systèmes répartis répliquant 
  des données de manière optimiste.
  Le paradigme de la réplication optimiste est que les sites composant
  le système réparti peuvent ré-éxecuter les requêtes des clients (actions)
  si la sémantique liant les actions le nécessite.
  Dans de tels systèmes le critère de cohérence est que les sites 
  convergent à terme vers des exécutions équivalentes.
  Afin d'assurer cette convergence, un protocole de validation est  nécessaire.
  C'est l'objet de cette étude.
  Notre protocole procède par éléctions successives sur des ensembles 
  d'actions exécutées de manière optimiste par le système.
  La sémantique prise en compte 
  dans ce protocole est suffisament riche pour exprimer des notions
  telles que la non-commutativité, le conflit ou encore la causalité 
  entre les actions.
  Nous prouvons que notre protocole est sûr, et ce en dépit des 
  éventuelles pannes franches pouvant survenir sur les sites.
  }
\RRabstract{
  \noindent
  We study large-scale distributed cooperative systems that use optimistic
  replication.
  We represent a system as a graph of actions (operations) connected by
  edges that reify semantic constraints between actions.
  Constraint types include conflict, execution order, dependence,
  and atomicity.
  The local state is some schedule that conforms to the constraints;
  because of conflicts, client state is only tentative.
  For consistency, site schedules should converge; we designed a
  decentralised, asynchronous commitment protocol.
  Each client makes a proposal, reflecting its tentative and{\slash}or
  preferred schedules.
  Our protocol distributes the proposals, which it decomposes into
  semantically-meaningful units called candidates, and runs an election
  between comparable candidates.
  A candidate wins when it receives a majority or a plurality.
  The protocol is fully asynchronous: each site executes its tentative
  schedule independently, and determines locally when a candidate has won
  an election.
  The committed schedule is as close as possible to the preferences
  expressed by clients.
}
\RRmotcle
    {
      réplication optimiste, validation, protocoles de vote
    }
\RRkeyword
    {
      data replication, optimistic replication, semantic
      replication, commitment, voting protocols.
      
    }
\RRprojet{Regal}
\RRtheme{\THCom}
\URRocq 
\begin{document}
\makeRR

\section{Introduction}

In a large-scale cooperative system, access to shared data is a
performance and availability bottleneck.
%%% This problem will only get worse as more mutable data is shared
%%% remotely, and as the gap between processing speeds and
%%% memory{\slash}network latency continues to widen.
One solution is optimistic replication (OR), where a process may read or
update its local replica without synchronising with remote sites
\cite{rep:syn:1500}.
OR decouples data access from network access.

In OR, each site makes progress independently, even while others are
slow, currently disconnected, or currently working in
isolated mode.
OR is well suited to peer-to-peer systems and to devices with occasional
connectivity.

%%% OR insulates users from network disruption, and may improve network
%%% utilisation by batching.
%%% OR supports mobile computers with slow, expensive or intermittent network
%%% connections, and wide-area networks with high and variable latencies.
%%% OR is especially useful for loosely-coupled co-operative work, where
%%% each user works on a separate copy, and synchronises only occasionally
%%% with co-workers.

Some limited knowledge of semantics provides a lot of extra power and
flexibility.
Therefore, we model the system as a graph, called a multilog, where each
vertex represents an action (i.e., an operation proposed by some client),
and an edge is a semantic relation between vertices, called a
constraint.
Our constraints include conflict, ordered execution, causal dependence,
and atomicity.
Each site has its own multilog, which contains actions submitted by the
local client, and their constraints, as well as those received from
other sites.
The current state is some execution schedule that contains actions from
the site's multilog, arranged to conform with its constraints.
For instance, when actions are antagonistic, at least one must abort; an
action that depends on an aborted action must abort too; non-commutative
actions should be scheduled in the same order everywhere, etc.
The site may choose any conforming schedule, e.g., one that minimises
aborts, or one that reflects user preferences.

For consistency, sites should agree on a common, stable and correct
schedule.
We call this agreement \emph{commitment}.
Some cooperative OR systems never commit, such as Roam
\cite{sys:rep:1585} or {D}raw-{T}ogether \cite{alg:app:1584}.
Previous work on commitment for semantic OR such as Bayou
\cite{syn:rep:1433} or IceCube \cite{syn:1474} centralises the agreement
at a central site.
Other work decentralises commitment (e.g., Paxos
consensus \cite{syn:rep:1548})
but ignores semantics.
It is difficult to reconcile semantics and decentralisation.
One possible approach would use Paxos to compute a total order, and abort
any actions for which this order would violate a constraint.
However this approach aborts actions unnecessarily.
Furthermore, the arbitrary total order may be very different from what
users expect.

A better approach is to order only non-commuting pairs of actions, to
abort only when actions are antagonistic, to
minimise dependent aborts, and to remain close to user expectations.
We propose an efficient, decentralised protocol that uses semantic
information for this purpose.
Participating sites make and exchange proposals asynchronously; our
algorithm decomposes each one into semantically-meaningful candidates;
it runs elections between comparable candidates.
A candidate that collects a majority or a plurality wins its election.
Voting ensures that the common schedule is similar to the tentative
schedules, minimising user surprise.
Our protocol orders only non-commuting actions and minimises unnecessary
aborts.

This paper makes several contributions:
\vspace{-.5ex}
\begin{itemize}
  \item
        Our algorithm combines a number of known techniques in a novel
        manner.
  \item
        We identify the concept of a semantically-meaningful unit for
        election (which we call a candidate).
  \item
        We propose an efficient commitment protocol system that is both
        decentralised and semantic-oriented, and that has weak communication
        requirements.
  \item
        We show how to minimise user surprise, the committed schedule
        being similar to local tentative schedules.
  \item
        We prove that the protocol is safe
        even in the presence of non-byzantine faults.
        The protocol is live as long as a sufficient number of votes are
        received.
\end{itemize}
\vspace{-.5ex}

The outline of this paper follows.
Section~\ref{sec:system-model} introduces our system model and our
vocabulary.
Section~\ref{sec:classical} discusses an abstraction of classical OR
approaches that is later re-used in our algorithm.
Section~\ref{sec:client} specifies client behaviour.
Our commitment protocol is specified in Section~\ref{sec:algorithms}.
Section~\ref{sec:proofoutline} provides a proof outline and adresses
message cost.
We compare with related work in
Section~\ref{sec:related}.
In conclusion, Section~\ref{sec:conclusion} discusses our results and
future work.

\section{System model}
\label{sec:system-model}

Following the ACF model \cite{rep:syn:1484}, an OR system is an
asynchronous distributed system of $n$ \emph{sites} $i, j, \ldots \in
\Sites$.
A site that crashes eventually recovers with its identity and persistent
memory intact (but may miss some messages in the interval).
Clients propose actions (deterministic operations) noted $\alpha, \beta,
\ldots{} \in A$.
An action might request, for instance, ``Debit $100$ euros from bank
account number 12345.''

A \emph{multilog} is a quadruple \Mmultilog, representing a graph where
the vertices
$K$ are actions, and \NotAfter, \Enables and \NonCommuting
(pronounced Not\-After, Enables and Non\-Com\-mut\-ing respectively) are
three sets of edges called \emph{constraints}.
We will explain their semantics shortly.%
\footnote{
    Multilog union, inclusion, difference, etc., are defined as
    component-wise union, inclusion, difference, etc., respectively.
    For instance if \Mmultilog and
    $M'=(K',\NotAfterPrime,\EnablesPrime,\NonCommutingPrime)$ their union is
    $M \union M' = (
    K \union K',
    \NotAfter \union \NotAfterPrime,
    \Enables \union \EnablesPrime,
    \NonCommuting \union \NonCommutingPrime)$.
    }

We identify a state with a \emph{schedule} $S$, a sequence of distinct
actions ordered by $<_{S}$ executed from the common initial state
$\init$.
The following safety condition defines semantics of Not\-After and
Enables in relation to schedules.
We define \soundM, the set of schedules $S$ that are \emph{sound} with
respect to multilog $M$, as follows:
\begin{displaymath}%%\onehalfspacing
  S \in \sound{M} \equaldef
  \Forall{\alpha, \beta \in \A}
  \left\lbrace
  \begin{array}{l}
    \init \in S \\
    \alpha \in S \implies \alpha \in K \\
    \alpha \in S \land \alpha \neq \init \implies \init <_{S} \alpha \\
    (\NotAfterAB) \land \alpha, \beta \in S \implies \alpha <_{S} \beta \\
    (\EnablesAB) \implies (\beta \in S \implies \alpha \in S ) \\
  \end{array}
\right.
\end{displaymath}

\noindent
Constraints represent scheduling relations between actions: Not\-After
is a (non-transitive) ordering relation and Enables is right-to-left
(non-transitive) implication.%
\footnote{
    A constraint is a relation in $A\times A$.
    By abuse of notation, for some relation $\mathcal{R}$,
    we write equivalently $(\alpha
    \mathrel{\mathcal{R}} \beta) \in M$ or $\alpha
    \mathrel{\mathcal{R}} \beta$ or $(\alpha, \beta) \in \mathcal{R}$.
    \NonCommuting is symmetric and \Enables is reflexive.
    They do not have any further special properties; in particular,
    \NotAfter and \Enables are not transitive, are not orders, and may be
    cyclic.
}
%

%%% Marc: NotAfter is non-transitive by design!  Otherwise
%%% NotAfter cycles would not be allowed.
%%% Enables is also
%%% non-transitive by design (although the reason is less compelling).
%%% Non-transitive Enables is a simplification.  If it was transitive,
%%% it would force the code to compute the transitive closure every time
%%% it adds an Enables to its multilog.  Transitive Enables forced me to
%%% invent the special Split-MustHave to support partial replication.
%%% If Enables is not transitive these problems go away.

Constraints represent semantic relations between actions.
For instance, consider a database system (more precisely, a serialisable
database that transmits transactions by value, such as DBSM
\cite{db:rep:1586}).
Assume shared variables $x, y, z$ are initially zero.
Two concurrent transactions $T_{1} = r(x)0; w(z)1$ and $T_{2} = w(x)2$ are
related by \NotAfterPair{T_{1}}{T_{2}}, since $T_{1}$ read a value that precedes
$T_{2}$'s write.%
\footnote{
    $r(x)n$ stands for a read of $x$ returning
    value $n$, and $w(x)n$ writes the value $n$ into $x$.
}
$T_{1}$ and $T_{3} = r(z)0; w(x)3$ are antagonistic, i.e., one or the other (or
both) must abort, as each is Not\-After the other.
In the execution $T_{1}; T_{4}$ where $T_{4} = r(z)1$, the latter transaction
depends causally on the former, i.e., they may run only in that order,
and $T_{4}$ aborts if $T_{1}$ aborts; we write \CausePair{T_{1}}{T_{4}}.
As another example,  Section~\ref{sec:application} discusses how to
encode the semantics of database transactions with constraints.

Non-commutativity imposes a liveness obligation: the system must put a
Not\-After between non-commuting actions, or abort one of them.
(Therefore, non-commutativity does not appear in the above safety
condition.)
The system also has the obligation to resolve antagonisms by
aborting actions.

For instance, transactions $T_{1}$ and $T_{5} = r(y)0$ commute if $x, y$ and
$z$ are independent.
In a database system that commits operations (as opposed to commiting
values), transactions $T_{6} = $``Credit 66 euros to Account 12345'' and
$T_{7} = $``Credit 77 euros to Account 12345'' commute since addition is a
commutative operation, but $T_{6}$ and $T_{8} = $``Debit 88 euros from Account
12345'' do not, if bank accounts are not allowed to become negative.
We write \NonCommutingPair{T_{6}}{T_{8}}.

Order, antagonism and non-commutativity are collectively called
conflicts.%
\footnote{
    Some authors suggest to remove conflicts by transforming the actions
    \cite{alg:rep:1460}.
    We assume that, if such transformations are possible, they have
    already been applied.
}

Clients submit actions to their local site; sites exchange actions
and constraints asynchronously.
The current knowledge of Site~$i$ at time $t$ is the distinguished
\emph{site-multilog} $\Mit$.
Initially, $\MIdxTime{i}{0} = \emptymultilog$, and it grows over time,
as we will explain later.
A site's current state is the \emph{site-schedule} $\Sit$, which is
some (arbitrary) schedule $\in \soundMit$.

An action executes tentatively only, because of conflicts and related
issues.
However, an action might have sufficient constraints that its execution is
stable.
We distinguish the following interesting subsets of actions relative to $M$.
  \begin{itemize}
    \item \emph{Guaranteed} actions appear in every
      schedule of \soundM.
      Formally, \guarM is the smallest subset of $K$ satisfying:
      $
      \init \in \guarM
      \land
      (( \alpha \in \guarM 
         \land
         \EnablesBA
        ) \implies \beta \in \guarM )
      $.
%%%       $
%%%       \{ \init \} \union
%%%       \{ \alpha \in A | \ExistsSuchthat{\beta \in \guarM}{\EnablesAB} \}$

    \item \emph{Dead} actions never appear in a schedule of \soundM.
%%%       \deadM is the greatest subset of $A$ containing
%%%       $
%%%       \{ \alpha \in A |
%%%          \ExistsSuchthat{\beta_{1}, \ldots{}, \beta_{m \geq 0} \in \guarM}
%%%                         {\alpha \NotAfter \beta_{1} \NotAfter \ldots{}
%%%                              \NotAfter \beta_{m} \NotAfter \alpha}
%%%       \}$
%%%       $
%%%       \union~~
%%%       \{ \alpha \in A |
%%%       \ExistsSuchthat{\beta \in \deadM}{\EnablesBA} \}
%%%       $
       \deadM is the smallest subset of $A$ satisfying:
       $
       ( (\alpha_{1}, \ldots{}, \alpha_{m \geq 0} \in \guarM)
          \land
          (\beta \NotAfter \alpha_{1} \NotAfter \ldots{}
                              \NotAfter \alpha_{m} \NotAfter \beta
         ) \implies \beta \in \deadM
       )
       \land
       ( ( \alpha \in \deadM
           \land
           \EnablesAB
          ) \implies \beta \in \deadM )
      $.

    \item \emph{Serialised} actions are either dead or ordered with respect
      to all non-commuting constraints.
      $\serialised{M} \equaldef
      \{ \alpha \in K |
      \Forall{\beta \in K,} \NonCommutingAB \implies
      \NotAfterAB \lor
      \NotAfterBA \lor
      \beta \in \dead{M} \lor
      \alpha \in \dead{M}
      \}
      $.

    \item \emph{Decided} actions are either dead, or
      both guaranteed and serialised.
      $\decided{M} \equaldef
      \deadM
      \union
      ( \guarM \inter \serialised{M})
      $.

    \item \emph{Stable} (i.e., \emph{durable}) actions are
          decided, and all actions that precede them by Not\-After or Enables
          are themselves stable:
          $\stableM \equaldef \deadM \union \{ \alpha \in \guarM \inter
          \serialised{M} |
          \Forall{\beta \in A} (\NotAfterBA \lor \EnablesBA) \implies \beta \in \stableM \}$.%

  \end{itemize}
\noindent
%%% These sets grow as \M grows.

To \emph{decide} an action \Alpha relative to a multilog $M$, means to add
constraints to the $M$, such that $\alpha \in \decidedM$.
In particular, to guarantee \Alpha, we add \EnablesAI to the multilog,
and to kill \Alpha, we add \NotAfterAA; to serialise non-commuting
actions \Alpha and \Beta, we add either \NotAfterAB, \NotAfterBA,
\NotAfterAA, or \NotAfterBB.

Multilog \M is said {sound} iff $\sound{M} \neq \emptyset$, or
equivalently, iff $\deadM \inter \guarM = \emptyset$.
An unsound multilog is definitely broken, i.e., no possible schedule can
satisfy all the constraints, not even the empty schedule.

Referring to the standard database terminology, a committed action is
one that is both stable and guaranteed, and aborted is the same as dead.
 
%%% \subsection{Correctness condition}

The standard correctness condition in OR systems is Eventual
Consistency: if clients stop submitting, eventually all sites reach the
same state.
We extend this definition by not requiring that clients stop, by
requiring that all states be correct, and by demanding decision.
%%% Pierre: i don't see in the follwoing definition that we exclude
%%% trivial solution %%%
%%% Marc: The liveness conditions ensure the system executes schedules
%%% (although I don't know how to exclude really bad schedules).

\begin{definition}{Eventual Consistency.}
  \label{def:eventual-consistency}
  An OR system is eventually consistent iff it satisfies all the following
  conditions:
  \begin{itemize}
  \item Local soundness (safety): Every site-schedule is sound:
        $\Forall{i, t} \Sit \in \soundMit$
  \item Mergeability (safety): The union of all the site-multilogs over
        time is sound:
        \begin{displaymath}
          \sound{\bigunion\limits_{i,t} \Mit} \neq \emptyset
        \end{displaymath}
    \item Eventual propagation (liveness):
        $\Forall{i,j \in \Sites}
         \Forall{t}
         \ExistsSuchthat{t'}{\Mit \subseteq M_{j}(t')}
         $
    \item Eventual decision (liveness):
          Every submitted action is eventually decided:
          \begin{displaymath}
            \Forall{\alpha \in A}
            \Forall{i \in \Sites}
            \Forall{t}
            \ExistsSuchthat{t'}{K_{i}(t) \subseteq \decided{M_i(t')}}
          \end{displaymath}
  \end{itemize}
\end{definition}

%%% Local soundness guarantees that every execution is safe.
%%% Mergeability ensures that globally, the system remains sound.
%%% Eventual propagation ensures that information available at some site is
%%% eventually known everywhere.
%%% Eventual decision guarantees that every action is eventually decided.

\begin{figure}[tbp]
  \begin{center}
  \begin{tabular}{ccccc|cc}
 ~~\Alpha~~& ~~$<$~~~  & ~~\Beta~~  & ~~$<$~~~ & ~~\Gamma~~   & \multicolumn{2}{c}{~~Decision~~} \\
    \hline
           &           & \Beta & \NonCommuting & \Gamma & (Serialise)     & \NotAfterBC \\
     guar. & $\BackwardsNotAfterSymbol$ & \Beta&&       & (Kill \Beta)    & \NotAfterBB  \\
     dead  & \Enables  & \Beta &               &        & (\Beta is dead) & \\
           &           & \Beta & $\BackwardsEnablesSymbol$ & \Gamma & (Kill \Beta) & \NotAfterBB \\
     \hline
     \multicolumn{5}{c|}{~\Beta not dead by above rules~} & (Guarantee \Beta) & \EnablesBI \\
  \end{tabular}
  \end{center}
\caption{\ProposalConservative: Applying semantic constraints to a given total order}
\label{fig:constraints+order}
\end{figure}

We assume some form of epidemic communication to fulfill Eventual
Propagation.
A commitment algorithm aims to fulfill the obligations of Eventual
Decision.
Of course, it must also satisfy the safety
requirements.

\section{Classical OR commitment algorithms}
\label{sec:classical}

Our proposal builds upon existing commitment algorithms for OR systems.
Generally, these either are centralised or do not take constraints into
account.
We note \proposalalgo{M} some algorithm that offers decisions based on
multilog \M; with no loss of generality, we focus on the outcome of
\Proposalalgo at a single site.
Assuming $M$ is sound, and noting the result $M' = \proposalalgo{M}$,
\Proposalalgo must satisfy these requirements:
\begin{itemize}%\onehalfspacing
  \item \Proposalalgo extends its input: $M \subseteq M'$.
  \item \Proposalalgo may not add actions: $K' = K$.
  \item \Proposalalgo may add constraints, which are restricted to decisions:
    \begin{center}
    $
    \begin{array}{rcl}
      \NotAfterPrimeAB &\implies& (\NotAfterAB) \lor (\NonCommutingAB) \lor (\beta = \alpha) \\
      \EnablesPrimeAB &\implies& (\EnablesAB) \lor (\beta = \init)\\
      \NonCommutingPrime &=& \NonCommuting
    \end{array}
    $
    \end{center}
  \item $M'$ is sound.
  \item $M'$ is stable: $\stable{M'}=K$.
    %% what follows does not work with our algorithm, M' must be entirely
    %% decided
    %%  If invoked sufficiently often, \Proposalalgo eventually decides:
    %%         For any non-decreasing series of sound multilogs $M^{0}
    %%         \subseteq M^{1} \subseteq
    %%         \ldots{} \subseteq M^{k} \subseteq \ldots{}$
    %%  we have:
    %%         $\Forall{i}, \Forall{\alpha \in K^{i}}
    %%          \ExistsSuchthat{j}
    %%                         {\alpha \in \decided{\proposalalgo{M^{j}}}}
    %%         $.
\end{itemize}

\noindent
\Proposalalgo could be any algorithm satisfying the requirements.

One possible algorithm, \ProposalConservative, first orders actions,
then kills actions for which the order is unsafe.
It proceeds as follows (see Figure~\ref{fig:constraints+order}).
Let $<$ be a total order of actions and $M$ a sound multilog.
The algorithm decides one action at at time, varying over all actions,
left to right; call the current action \Beta.
Consider actions \Alpha and \Gamma such that $\alpha < \beta < \gamma$:
\Alpha has already been decided, and \Gamma has not.
If \NonCommutingBC, then serialise them in schedule order.
If \NotAfterBA, and \Alpha is guaranteed, kill \Beta, because the
schedule and the constraint are incompatible.
If \EnablesCB, conservatively kill \Beta, because it is not known
whether \Gamma can be guaranteed.
By definition, if \EnablesAB and \Alpha is dead, then \Beta is dead.
If \Beta is not dead by any of the above rules, then decide \Beta
guaranteed (by adding \EnablesBI to the multilog).
The resulting $\sound{\ProposalConservative(M)}$ contains a unique
schedule.

It should be clear that this approach is safe but tends to
kill actions unnecessarily.

The Bayou system \cite{syn:rep:1433} applies \ProposalConservative,
where $<$ is the order in which actions are received at a single primary
site.
An action aborts if it fails an application-specific precondition, which
we reify as a \NotAfter constraint.

In the Last-Writer-Wins (LWW) approach \cite{bd:rep:1454}, an action
(completely overwriting some datum) is stamped with the time it is
submitted.
Two actions that modify the same datum are related by \NotAfter in
timestamp order.
Sites execute actions in arbitrary order and apply \ProposalConservative.
Consequently, a datum has the state of the most recent write (in
timestamp order).

%%% Alternatively, \ProposalRandom, defined along the following lines, gives good
%%% results \cite{formel:rep:1529}:
%%% (1) Choose some arbitrary Non\-Com\-mut\-ing relation \NonCommutingAB; decide
%%%     \NotAfterAB.  Iterate until all Non\-Com\-mut\-ing relations are serialised.
%%% (2) Choose some Not\-After cycle; decide an arbitrary one of its actions dead.
%%%     Iterate until all cycles contain at least one dead action.
%%% (3) For all Enables constraints \EnablesAB: if \Alpha is dead, mark
%%%     \Beta dead.
%%% (4) Decide all remaining actions guaranteed.

The decisions computed by the above systems are mostly arbitrary.
A better way would be to minimise aborts, or to follow user
preferences, or both.
This was the approach of the IceCube system \cite{syn:1474}.
\ProposalIceCube is an optimization algorithm that
minimises the number of dead actions in $\ProposalIceCube(M)$.
It does so by heuristically comparing all possible sound schedules that
can be generated from the current site-multilog.
The system suggests a number of possible decisions to the user, who
states his preference.

Except for LWW, which is decentralised but deterministic, the above
algorithms centralise commitment at a primary site.
%%% Bayou, \ProposalRandom, and IceCube are non-deterministic;
%%% therefore commitment is centralised at a primary site.
%%% In LWW, timestamp order is deterministic, and actions overwrite their
%%% datum entirely, therefore it is safe for every site to execute
%%% \ProposalConservative despite non-deterministic local schedules.

To decentralise decision, one approach might be to determine a global
total order $<$, using a decentralised consensus algorithm such as Paxos
\cite{syn:rep:1548}, and apply \ProposalConservative.
As above, this order is arbitrary and \ProposalConservative tends to
kill unnecessary.
Instead, our algorithm allows each site to propose decisions that
minimises aborts and follows local client preferences, and to reach
consensus on these proposals in a decentralised manner.
This is the subject of the rest of this paper.

\section{Client operation}
\label{sec:client}

We now begin the discussion of our algorithm.
We start with a specification of client behaviour.

\begin{algorithm}[tbp]
  \caption{\ClientActionsConstraints{L}}
  \label{alg:client}
  \begin{algorithmic}[1]
    \REQUIRE $L \subseteq A$
    \STATE $\Ki \assign \Ki \union L$
    \FOR{all $(\alpha, \beta) \in \Ki \times \Ki$
         such that
            $\NotAfterIdxPair{\CM}{\alpha}{\beta}$}
         \STATE $\NotAfteri \assign \NotAfteri \union
                                      \{ ( \alpha, \beta ) \}$
    \ENDFOR
    \FOR{all $(\alpha, \beta) \in \Ki \times \Ki$
         such that
            $\EnablesIdxPair{\CM}{\alpha}{\beta}$}
         \STATE $\Enablesi \assign \Enablesi \union
                                      \{ ( \alpha, \beta ) \}$
    \ENDFOR
    \FOR{all $(\alpha, \beta) \in \Ki \times \Ki$
         such that
            $\NonCommutingIdxPair{\CM}{\alpha}{\beta}$}
         \STATE $\NonCommutingi \assign \NonCommutingi \union
                                      \{ ( \alpha, \beta ) \}$
    \ENDFOR
  \end{algorithmic}
\end{algorithm}

\subsection{Client Behaviour and client interaction}
\label{sec:transition-rules}

An application performs tentative operations by submitting actions
and constraints to its local site-multilog;
they will eventually propagate to all sites.

%% The site-multilog also contains constraints that reflect application
%% semantics.
%% For instance, if some action \Alpha computes a value, and a later action
%% \Beta depends causally on that value, this should be reflected in the
%% dependency \CauseAB.
%% Similarly, if a group of actions $\{ \alpha, \beta, \gamma \}$ has
%% all-or-nothing semantics, this is reflected by \Enables cycle $ \alpha
%% \Enables\beta \Enables \gamma \Enables \alpha$.
%% If the application has submitted action \Alpha, and receives remote
%% action \Beta that is non-commuting (resp.~antagonistic) with it, the
%% constraint \NonCommutingAB (resp.~\AntagonismAB) should appear in the
%% multilog.

We abstract application semantics by postulating that clients have
access to a sound multilog containing all the semantic constraints: $\CM =
\multilog{A}{\NotAfter_{\CM}}{\Enables_{\CM}}{\NonCommuting_{\CM}}$.
For an example \CM, see Section~\ref{sec:application}.

As the client submits actions $L$ to the
site-multilog, function $\mathit{ClientActionsConstraints}$
(Algorithm~\ref{alg:client}) adds constraints with respect to actions
that the site already knows.%
\footnote{
    In the pseudo-code, we leave the current time $t$ implicit.
    A double-slash and sans-serif font indicates a comment, as  in
    \algorithmiccomment{This is a comment}.
    }

To illustrate, consider Alice and Bob working together.
Alice uses their shared calendar at \siteone, and Bob at \sitetwo.
Planning a meeting with Bob in Paris, Alice submits two actions:
$\Alpha=$``Buy train ticket to Paris next Monday at 10:00''
and
$\Beta=$``Attend meeting''.
As \Beta depends causally on \Alpha, \CM contains
$\NotAfterIdxPair{\CM}{\alpha}{\beta} \land \EnablesIdxPair{\CM}{\alpha}{\beta}$.
Alice calls \ClientActionsConstraints{\{\alpha\}} to add action \Alpha
to site-multilog \MIdx{1}, and, some time later, similarly for \Beta.
At this point, Algorithm~\ref{alg:client} adds the constraints
\NotAfterPair{\alpha}{\beta} and
\EnablesPair{\alpha}{\beta} taken from \CM.

\subsection{Multilog Propagation}
\label{sec:multilogProp}

When a client adds new actions $L$ into a site-multilog, $L$ 
and the constraints computed by $\mathit{ClientActionsConstraints}$,
form a multilog that is sent to remote sites.
Upon reception, receivers merge this multilog into their own site-multilog.
By this so-called epidemic communication \cite{alg:rep:1459}, every site
eventually receive all actions and constraints submitted at any site.

When \sitei receives a multilog $M$, it executes function
$\mathit{ReceiveAndCompare}$ (Algorithm~\ref{alg:conflict}), which
first merges what it received into the local site-multilog.
Then, if any conflicts exist between previously-known
actions and the received ones, it adds the corresponding constraints to
the site-multilog.%
\footnote{
    $\mathit{ClientActionsConstraints}$ provides constraints between
    successive actions submitted at the same site.
    These consist typically of dependence and atomicity constraints.
    In contrast, $\mathit{ReceiveAndCompare}$ computes constraints
    between independently-submitted actions.
}

\begin{algorithm}[!tbp]
  \caption{\ReceiveAndCompare{M}}
  \label{alg:conflict}
  \begin{algorithmic}
    \DECLARE \Mmultilog a multilog receives from a remote site
       \STATE $\Mi \assign \Mi \union M$

    \FOR{all $(\alpha, \beta) \in \Ki \times \Ki$
         such that
            $\NotAfterIdxPair{\CM}{\alpha}{\beta}$}
         \STATE $\NotAfteri \assign \NotAfteri \union
                                      \{ ( \alpha, \beta ) \}$
    \ENDFOR
    \FOR{all $(\alpha, \beta) \in \Ki \times \Ki$
         such that
            $\NonCommutingIdxPair{\CM}{\alpha}{\beta}$}
         \STATE $\NonCommutingi \assign \NonCommutingi \union
                                      \{ ( \alpha, \beta ) \}$
    \ENDFOR
  \end{algorithmic}
\end{algorithm}

%JPB:I removed this because it is out of our scope (we are not concerned with
%upd.prop.)
%For instance, every site sends its site-multilog to some arbitrary other
%site at some arbitrary interval.
%(In practice, it only needs to send information the receiver is missing.)
%The receiver adds the data received into its own site-multilog.
%If the communication graph is sufficiently dense, and sites communicate
%sufficiently often, then every action or constraint added at some site
%eventually reaches every other site with high probability \cite{???}.

%% PIERRE : this was unsound with the timestamping of proposals.
% I had the choice between suppressing timestamping and 
% considering the communication atomic, or suppressing atomic communication
% and changin the way multilog are sent. I chose the later.

% To simplify exposition, we will assume here that communication is
% all-or-nothing: if communication succeeds, the receiver receives the
% full state of the sender's multilog.
% The protocol remains correct under weaker, FIFO-like assumptions.

Let us return to Alice and Bob.
Suppose that Bob now adds action \Gamma, meaning ``Cancel the meeting,''
to \MIdx{2}.
Action \Gamma is antagonistic with action \Beta; hence,
\AntagonismIdxPair{\CM}{\beta}{\gamma}.
Some time later, \sitetwo sends its site-multilog to \siteone; when \siteone
receives it, it runs Algorithm~\ref{alg:conflict},
notices the antagonism, and adds constraint
\AntagonismPair{\beta}{\gamma} to \MIdx{1}.
Thereafter, site-schedules at \siteone may include either \Beta or
\Gamma, but not both.

\section{A decentralised commitment protocol}
\label{sec:algorithms}

Epidemic communication ensures that all site-multilogs eventually
receive all information, but site-schedules might still differ between
sites.

For instance, let us return to Alice and Bob.
Assuming users add no more actions, eventually all site-multilogs
become \multilog{ \{ \init, \alpha, \beta, \gamma \} }
                 { \{ \NotAfterAB, \NotAfterPair{\beta}{\gamma}, \NotAfterPair{\gamma}{\beta} \} }
                 { \{ \EnablesAB \} }
                 { \emptyset }.
In this state, actions remain tentative; at
time $t$, \siteone might execute $\siteschedule{1}{t}=\init;\alpha;\beta$, \sitetwo
$\siteschedule{2}{t}=\init;\alpha;\gamma$, and just $\init$ at $t+1$.
A commitment protocol ensures that \Alpha, \Beta and \Gamma eventually
stabilise, and that both Alice and Bob learn the same outcome.
For instance, the protocol might add \EnablesPair{\beta}{\init} to
\MIdx{1}, which guarantees \Beta, thereby both guaranteeing \Alpha and
killing \Gamma.
\Alpha, \Beta and \Gamma are now decided and stable at \siteone.
\MIdx{1} eventually propagates to other sites; and inevitably, all
site-schedules eventually start with $\init; \alpha; \beta$, and \Gamma is
dead everywhere.

\subsection{Overview}

Our key insight is that eventual consistency is equivalent to the
property that the site-multilogs of all sites share a common
\emph{well-formed prefix} (defined hereafter) of stable actions, which
grows to include every action eventually.
Commitment serves to agree on an extension of this prefix.
As clients continue to make optimistic progress beyond this prefix, the
commitment protocol can run asynchronously in the background.

In our protocol, different sites run instances of \Proposalalgo to make
proposals; a proposal being a tentative well-formed prefix of its
site-multilog.
Sites agree via a decentralised election.
This works even if \Proposalalgo is non-deterministic, or if sites use
different \Proposalalgo algorithms.
We recommend IceCube \cite{syn:1474} but any algorithm satisfying the
requirements of Section~\ref{sec:classical} is suitable.

%%% They participate in several concurrent elections at any point in time;
%%% however agents do not synchronize their elections, instead they vote
%%% asynchronously and continuously on all current candidates.

%%% A site may not retract any part of a proposal until it either wins or
%%% loses the corresponding election.
%%% No acceptor may elect incompatible candidates; this ensures that every
%%% site-multilog remains sound.
%%% Together with the transition rules, this ensures mergeability.

In what follows, $i$ represents the current site, and $j, k$ range over
$\Sites$.

We distinguish two roles at each site, proposers and acceptors.
Each proposer has a fixed \emph{weight}, such that
$\sum_{k \in \Sites}\weightk = 1$.
In practice, we expect only a small number of sites to have non-zero
weights (in the limit one site might have weight 1, this is a primary
site as in Section~\ref{sec:classical}), but the safety of our
protocol does not depend on how weights are allocated.
To simplify exposition, weights are distributed ahead of time and do not
change; it is relatively straightforward to extend the current
algorithm, allowing weights to vary between successive elections.

An acceptor at some site computes the outcome of an election, and
inserts the corresponding decision constraints into the local
site-multilog.

Each site stores the most recent proposal received from each proposer
in array $proposals_i$,  of size $n$ (the number of sites).
To keep track of proposals, each entry $proposals_i[k]$ carries a logical
timestamp, noted $proposals_i[k].ts$. Timestamping ensures the liveness
of the election process despite since links between nodes are not 
necessarily FIFO.

 \begin{algorithm*}[!ht]
   \caption{Algorithm at \sitei}
   \label{alg:eca}
  \begin{algorithmic}[1]
    \DECLARE $\Mi$: local site-multilog
    \DECLARE $\proposalsi[n]$: array of proposals, indexed by
             site; a proposal is a multilog

    \STATE $\Mi \assign \emptymultilog$ \label{b:init}
    \STATE $\proposalsi \assign [ (\emptymultilog,0) , \ldots{},(\emptymultilog,0) ]$ \label{e:init}

    \LOOP [Epidemic transmission] \label{b:ep}
       \STATE Choose $j \neq i$;
       \STATE Send copy of \Mi and \proposalsi to $j$
    \ENDLOOP
    \STATE $||$
    \LOOP [Epidemic reception]
       \STATE Receive multilog \M and proposals $P$ from some site $j \neq i$
       \STATE \ReceiveAndCompare{M} \label{site:rcvcmp}
              \COMMENT{Compute conflict constraints}
       \STATE \MergeProposals{P}
    \ENDLOOP \label{e:ep}

    \STATE $||$

    \LOOP [Client submits]\label{b:client}
        \STATE Choose $L \subseteq A$
        \STATE \ClientActionsConstraints{L} \COMMENT{Submit actions, compute local constraints}
    \ENDLOOP \label{e:client}

    \STATE $||$

    \LOOP [Compute current local state] \label{b:sched}
       \STATE Choose $\Si \in \soundMi$
       \STATE Execute \Si
    \ENDLOOP \label{e:sched}

    \STATE $||$

    \LOOP [Proposer] \label{b:proposer}
       \STATE $\UpdateProposal$
              \COMMENT{Suppress redundant parts}
       \STATE $\proposalsi[i] \assign \proposalalgo{\Mi \union
              \proposalsi[i]}$
              \COMMENT{New proposal, keeping previous}
       \STATE \text{Increment } $\proposalsi[i].ts$
    \ENDLOOP \label{e:proposer}

    \STATE $||$

    \LOOP [Acceptor] \label{b:acceptor}
       \STATE \Elect
    \ENDLOOP \label{e:acceptor}

  \end{algorithmic}
\end{algorithm*}

Each site performs Algorithm~\ref{alg:eca}.
First it initialises the site-multilog and proposals data structures,
then it consists of a number of parallel iterative threads, detailed in
the next sections.
Within a thread, an iteration is atomic.
Iterations
are separated by arbitrary amounts of time.

\subsection{Epidemic communication}

The first two threads (lines~\Lines{ep}) exchange multilogs and
proposals between sites.
Function $\mathit{ReceiveAndCompare}$ (defined in Algorithm
\ref{alg:conflict}, Section \ref{sec:multilogProp}) compares actions
newly received to already-known ones, in order to compute conflict
constraints.
In Algorithm~\ref{alg:receive_proposals} a receiver updates its own set
of proposals with any more recent ones.

\subsection{Client, local state, proposer}

The third thread (lines~\Lines{client}) constitutes one half of the client.
An application submits tentative operations to
its local site-multilog, which the site-schedule will (hopefully)
execute in the fourth thread.
Constraints relating new actions to previous ones are included at this
stage by function $\mathit{ClientActionsConstraints}$ (defined in
Algorithm~\ref{alg:client}).

The other half of the client is function $\mathit{ReceiveAndCompare}$
(Algorithm~\ref{alg:conflict}) invoked in the second thread
(line~\ref{site:rcvcmp}).

The fourth thread (lines~\Lines{sched}) computes the current tentative
state by executing some sound site-schedule.
%%% It is possible that the current schedule does not linearly extend the
%%% previous one; this can be implemented as a roll-back followed by forward
%%% execution.

The fifth thread (\Lines{proposer}) computes proposals by invoking
\Proposalalgo.
A proposal extends the current site-multilog with proposed decisions.
A proposer may not retract a proposal that was already received by some
other site.
Passing argument $\Mi \union \proposalsi[i]$ to \Proposalalgo ensures
that these two conditions are satisfied.

However, once a candidate has either won or lost an election, it becomes
redundant; $\mathit{UpdateProposal}$ removes it from the proposal
(Algorithm~\ref{alg:proposal_update}).

The last thread is described in the next section.

\begin{algorithm}[tbp]
  \caption{\UpdateProposal}
  \label{alg:proposal_update}

  \begin{algorithmic}[1]
    \STATE Let $P = \stdmultilogIdx{P} = \proposalsi[i]$
    \STATE $K_{P} \assign K_{P} \setminus \decided{\Mi}$
    \STATE $\NotAfterIdx{P} \assign
            \NotAfterIdx{P} \inter K_{P} \times K_{P}$
    \STATE $\EnablesIdx{P} \assign
            \EnablesIdx{P} \inter K_{P} \times K_{P}$
    \STATE $\NonCommutingIdx{P} \assign \emptyset$
    \STATE $\proposalsi[i] \assign P$
  \end{algorithmic}
\end{algorithm}

\subsection{Election}

The last thread (\Lines{acceptor}) conducts elections.
Several elections may be taking place at any point in time.
An acceptor is capable of determining locally the outcome of
elections.
A proposal can be decomposed into a set of eligible candidates.

\subsubsection{Eligible candidates}
\label{sec:candidates}

A candidate cannot be just any subset of a proposal.
Consider, for instance, proposal $P = \multilog%
    { \{\init,\alpha,\gamma \} }%
    { \{\NotAfterAC, \NotAfterCA, \NotAfterAA\} }%
    { \{ \EnablesPair{\gamma}{\init} \} }%
    { \emptyset }$, and some candidate
$X$ extracted from $P$.
If $X$ could contain \Gamma and not \Alpha, then we might guarantee \Gamma
without killing \Alpha, which would be incorrect.
According to this intuition, $X$ must be a \emph{well-formed prefix} of
$P$:

\begin{definition}{Well-formed prefix.}
  %\begin{onehalfspacing}
  Let \Mmultilog and \MmultilogPrime be two multilogs.
  $M'$ is a \emph{well-formed prefix} of \M, noted \prefix{M'}{M},
  if
  (i) it is a subset of \M,
  (ii) it is stable,
  (iii) it is left-closed for its actions,
  and
  (iv) it is closed for its constraints.
  \begin{displaymath}
    \prefix{M'}{M} \equaldef
    \left\lbrace
    \begin{array}{l}
          M' \subseteq M\\
          K' = \stable{M'}\\
          \Forall{\alpha, \beta \in A}
              \beta \in K' \implies
              \left\lbrace
                 \begin{array}{l}
                   \NotAfterAB \implies \NotAfterPrimeAB \\
                   \EnablesAB \implies \EnablesPrimeAB \\
                   \NonCommutingAB \implies \NonCommutingPrimeAB \\
                 \end{array}
          \right.
      \\
      \Forall{\alpha, \beta \in A}
              ( \NotAfterPrimeAB \lor
              \EnablesPrimeAB \lor
              \NonCommutingPrimeAB ) \implies \alpha, \beta \in K' \\
    \end{array}
    \right.
  \end{displaymath}
  %\end{onehalfspacing}
\end{definition}

\noindent
A well-formed prefix is a semantically-meaningful unit of proposal.
For instance, if a \NotAfter or \Enables cycle is present in \M, every
well-formed prefix either includes the whole cycle, or none of its
actions.

Unfortunately, because of concurrency and asynchronous communication, it
is possible that some sites know of a \NotAfter cycle and not others;
or more embarassingly, that sites know only parts of a cycle.
Therefore we also require the following property:
\begin{definition}{Eligible candidates.}
  An action is \emph{eligible} in set $L$ if all its predecessors by
  client Not\-After, Enables and Non\-Com\-mut\-ing relations are in $L$.
  A candidate multilog \M is eligible if all actions in $K$ are eligible
  in $K$:
  $
  \eligible{M} \equaldef
      \Forall{\alpha, \beta \in A \times K}
        (\NotAfterIdxPair{\CM}{\alpha}{\beta} \lor
        \NonCommutingIdxPair{\CM}{\alpha}{\beta} \lor
        \EnablesIdxPair{\CM}{\alpha}{\beta} )
          \implies \alpha \in K
  $.
\end{definition}

To compute eligibility precisely would require local access to the
distributed state, which is impossible.
Therefore acceptors must compute a safe approximation (i.e., false
negatives are allowed) of eligibility.
For instance, in the database example, a sufficient condition for
transaction $T$ to be eligible at \SiteId{i} is that all transactions
submitted (at any site) concurrently with $T$ are also known at
\SiteId{i}.
Indeed, all such transactions have gone through either
$\mathit{ClientActionsConstraints}$ or $\mathit{ReceiveAndCompare}$;
hence according to Table~\ref{table:its:db}, $T$ is eligible.

\subsubsection{Computation of votes}

We define a vote as a pair $( \mathit{weight}, \mathit{siteId} )$.
The comparison operator for votes breaks ties by comparing site
identifiers:
$(w,i) > (w', i') \equaldef w > w' \lor ( w = w' \land i > i')$.
Therefore, votes add up as follows:
$(w, i) + (w', i') \equaldef (w+w', \max(i,i'))$.
Candidates are \emph{compatible} if their union is sound:
$\compatible{M}{M'} \equaldef \sound{M \union M'} \neq \emptyset$.
The votes of compatible candidates add up; $\tally{X}$ computes the
total vote for some candidate $X$:
\begin{displaymath}%\onehalfspacing
  \tally{X} \equaldef \sum\limits_{k: \prefix{X}{\proposalsi[k]}} ( \weightk, k)
\end{displaymath}

An election pits some candidate against \emph{comparable} candidates
from all other sites.
Two multilogs are comparable if they contain the same set of
actions: $\comparable{M}{M'} \equaldef K = K'$.
The direct opponents of candidate $X$ in some election are comparable
candidates that $X$ does not prefix:
\begin{displaymath}
\opponents{X} \equaldef
\{B |
\ExistsSuchthat{k}{\prefix{B}{\proposalsi[k]}}
\land
\comparable{B}{X}
\land \nprefix{X}{B})
\}
\end{displaymath}
However, we must also count missing votes, i.e., the weights of sites whose
proposals do not yet include all actions in X.
Function $\cotally{X}$ adds these up:
\begin{displaymath}%\onehalfspacing
  \begin{array}{c}
  \cotally{X} \equaldef \sum\limits_{k: K_{X} \not\subseteq K_{\proposalsi[k]}} ( \weightk, k)
  \end{array}
\end{displaymath}

\begin{algorithm}[tbp]
  \caption{\Elect}
  \label{alg:elect}
  \begin{algorithmic}[1]
    \STATE Let $X$ be a multilog such that:\\
    $
      \begin{array}{@{~~~~}rl}
          & \ExistsSuchthat{k \in \Sites}{\prefix{X}{\proposalsi[k]}} \\
    \land & X \not \subseteq \Mi \\
    \land & \eligible{X} \\
    \land & \tally{X} > \max\limits_{B \in \opponents{X}}(\tally{B}) + \cotally{X}
      \end{array}
    $
    \IF{such an $X$ exists}
       \STATE Choose such an $X$
       \STATE $\Mi \assign \Mi \union X$
    \ENDIF
  \end{algorithmic}
\end{algorithm}
Algorithm~\ref{alg:elect} depicts the election algorithm.
A candidate is a well-formed prefix of some proposal.
We ignore already-elected candidates and we only consider eligible
ones.
A candidate wins its election if its tally is greater than the tally of
any direct opponent, plus its cotally.
Note that, as proposals are received, cotally tends towards 0, therefore
some candidate is eventually elected.
We merge the winner into the site-multilog.

%%  We now futher detail what we mean by eligible candidates.
%% As clients submit new actions, and sites serialize non-commuting
%% ones, cycles may be created.
%% Now if we elect candidates by majority, we may erronously commit a \NotAfter cycle
%% of actions.
%% To remdedy this issue, we consider a safe approximation of the
%% maximum size of cycle in \M being constituted by \NotAfter and
%% \NonCommuting constraints.
%% We note this approximation \cycleM.
%% A candidate $X$ is eligible on site $i$ only if a set of sites whose
%% weight is greater than $1 - \frac{1}{\cycleM}$ have voted on $X_{K}$:%
%% %
%% \footnote{
%%   We prune the case where there is no cycle in \M, by taking at least
%%   a weight of $1/2$.
%% }%
%% %
%% \begin{displaymath}\onehalfspacing
%%   \begin{array}{c}
%%     \eligible{X} \equaldef
%%     \sum\limits_{k: K_{X} \in\subseteq K_{\proposalsi[k]}} \weightk > \max(1-\frac{1}{\cycleM},1/2)
%%   \end{array}
%% \end{displaymath}
%% A candidate wins its election if its tally is greater than the tally of
%% any direct opponent, plus its cotally.
%% Note that as proposals make progress and sites stay correct,
%% cotally tends towards 0, therefore some candidate is eventually elected.
%% We merge the winner into the site-multilog.

\begin{algorithm}[tbp]
  \caption{\MergeProposals{P}}
  \label{alg:receive_proposals}
  \begin{algorithmic}[1]
    \FORALL{$k$}
        \IF{$\proposalsi[k].ts < P[k].ts$}
        \STATE $\proposalsi[k] \assign P[k]$
        \STATE $\proposalsi[k].ts \assign P[k].ts$
        \ENDIF
    \ENDFOR
  \end{algorithmic}
\end{algorithm}

\subsection{Example}

%%JOAO: The comment below is the previous text (Pierre's modification
%%of the example). I returned to the previous example, though.
%We return to our example.
%Recall that, once  Alice and Bob have submitted their actions, and
%\siteone and \sitetwo have exchanged site-multilogs, both site-multilogs
%are equal to \multilog{ \{ \init, \alpha, \beta \} }
%                 { \{ \NotAfterAB, \NotAfterAC, \NotAfterCA \} }
%                 { \{ \EnablesAB \} }
%                 { \emptyset }.
%Now Alice (at \siteone) proposes to guarantee \Alpha and \Beta, and to kill
%\Gamma: $proposals_1[1]=\MIdx{1} \union \{\EnablesPair{\beta}{\init}\}$.
%Meanwhile, Bob at \sitetwo proposes to guarantee \Gamma and kill \Alpha.
%As \EnablesAB, \Beta is also killed (Alice does not buy it ticket):
%$proposals_2[2]=\MIdx{2} \union \{\EnablesPair{\gamma}{\init},
%\NotAfterPair{\alpha}{\alpha}\}$.
%These proposals are incompatible; therefore the commitment protocol
%eventually agrees on at most one of them.

We return to our example.
Recall that, once  Alice and Bob have submitted their actions, and
\siteone and \sitetwo have exchanged site-multilogs, both site-multilogs
are equal to \multilog{ \{ \init, \alpha, \beta \} }
                 { \{ \NotAfterAB, \NotAfterAC, \NotAfterCA \} }
                 { \{ \EnablesAB \} }
                 { \emptyset }.
Now Alice (\siteone) proposes to guarantee \Alpha and \Beta, and to kill
\Gamma: $proposals_1[1]=\MIdx{1} \union \{\EnablesPair{\beta}{\init}\}$.
In the meanwhile, Bob at \sitetwo proposes to guarantee \Gamma and \Alpha, and
to kill \Beta:
$proposals_2[2]=\MIdx{2} \union \{\EnablesPair{\gamma}{\init},
\EnablesPair{\alpha}{\init}\}$.
These proposals are incompatible; therefore that the commitment protocol
will eventually agree on at most one of them.

Consider now a third site, \sitethree; assume that the three sites have
equal weight $\frac{1}{3}$.
Imagine that \sitethree receives \sitetwo's site-multilog and proposal,
and sends its own proposal that is identical to \siteone's.
Sometime later, \sitethree sends its proposal to \siteone.
At this point, \siteone has received all sites' proposals.
Now \siteone might run an election, considering a candidate $X$ equal to
$proposals_1[1]$.
$X$ is indeed a well-formed prefix of $proposals_1[1]$; now suppose
that $X$ is eligible as all sites have voted on $K_{X}$;
$\tally{X}=\frac{2}{3}$ is greater than that of $X$'s only opponent
($\tally{proposals_1[2]}=\frac{1}{3}$); and $\cotally{X}=0$.

Therefore, \siteone elects $X$ and merges $X$ into \MIdx{1}.
Any other site will either elect $X$ (or some compatible candidate) or
become aware of its election by epidemic transmission of \MIdx{1}.

\section{Discussion}
\label{sec:proofoutline}

\subsection{Safety proof outline}

Section~\ref{def:eventual-consistency} states our safety property, the
conjunction of mergeability and local soundness.
Clearly Algorithm~\ref{alg:eca} satisfies local soundness;
see lines \Lines{sched}.
We now outline a proof of mergeability.

We say that candidate $X$ is \emph{elected} in a run $r$ at time $t$, if
some acceptor $i$ executes Algorithm~\ref{alg:elect} in $r$ at $t$, and
elects a candidate $Y$ such that $\prefix{X}{Y}$.
Given a run $r$ of Algorithm~\ref{alg:eca},
we note $\Elected(r,t)$ the set of candidates elected
in $r$ up to time $t$ (inclusive),
and $\Elected(r)$ the set of candidates elected during $r$.
Observe that, since $\CM$ is sound, Algorithm~\ref{alg:eca}
satisfies mergeability in a run $r$ if and only if the acceptors elect a
sound set of candidates during $r$
(
$
\bigunion_{
  \begin{subarray}{l}
    X \in \Elected(r)
  \end{subarray}
}X
$
is sound
).

%%% %
%%% \footnote{
%%%   This is stricly stronger than pair-wise compatibility,
%%%   which is insufficient: consider for instance
%%%   a cycle of three or more \NotAfter relations.
%%% }%

Suppose, by contradiction, that during run $r$,
this set is unsound.
As \CM is sound, by \Proposalalgo candidates are sound.
Consequently there must exist an unsound set of candidates
$C \subseteq \Elected(r)$.
Let us now consider the following property:
\begin{definition}{Minimality.}\label{def:minimality}
  A multilog \M is said minimal iff:
  $
  \Forall{M' \subseteq M}
     \prefix{M'}{M} \implies M'=M
  $.
\end{definition}
As candidates are eligible, there must exist
two candidates $X$ and $X'$ in $C$ such that:
(i) $X$ and $X'$ are non-compatible,
and
(ii) $X$ and $X'$ are minimal.

We define the following notation.
  Let $i$ (resp.~$i'$) be the acceptor
  that elects $X$ (resp.~$X'$) in $r$.
  $t$ is the time where $i$ elects $X$ in $r$
  (resp.~$t'$ for $X'$ on $i'$).
  For a proposer $k$, $t_k$ (resp.~${t'}_k$) is
  the time at which it sent $proposals_i[k](t)$ to $i$
  (resp.~$proposals_{i'}[k](t')$ to $i'$).
  $Q$ (resp.~$Q'$) is the set of proposers that
  vote for $X$ at $t$ on $i$ (resp.~for $X'$ at $t'$ on $i'$);
  formally $Q = \{ k | \prefix{X}{proposals_i[k](t)} \}$
  and $Q' = \{ k | \prefix{X'}{proposals_{i'}[k](t')} \}$.

Hereafter, and without loss of generality, we suppose that:
(i) $t < t'$,
(ii) $X$ is the first candidate non-compatible with $X'$ elected in $r$,
and
(iii) $\Elected(r,t'-1)$ is sound.

Since $i'$ elects $X'$ at $t'$, at that time on \SiteId{i'}:
\begin{equation}  \label{equ:1}
  %\onehalfspacing
  \tally{X'} > \max\limits_{B \in \opponents{X'}}(\tally{B}) + \cotally{X'}
\end{equation}

Equation~\ref{equ:1} defines an upper bound for \tally{X} on $i$ at $t$,
as follows.
Consider some $k \in Q$.
If $t_k < {t'}_k$ then
  from Algorithm~\ref{alg:proposal_update},
  and the fact that $\Elected(r,t'-1)$ is sound,
  we know that \prefix{X}{proposals_{i'}[k](t')}.

If now $t_k > {t'}_k$,
  then as \tally{X'}, \opponents{X'} and \cotally{X'} define a partition
  of \Sites,
  either:
  \begin{enumerate}
  \item $k$ has not yet voted on $K_{X'}$ at $t'$ on $i'$ and
  its weight is counted in \cotally{X'}.
  \item Or, if its vote already includes $K_{X'}$, it is counted in \opponents{X'}
    as $X$ is the first candidate non-compatible with $X'$ elected in $r$,
    \prefix{X}{proposals_i[k](t)}, and $\neg~\compatible{X}{X'}$.
  \end{enumerate}
%%   Note that $k$ can be counted in \tally{X'} is impossible:
%%   $X$ and $X'$ are not compatible, $X'$ and $X$ are minimal,
%%   and $\Elected(r,t-1)$ is sound.

From these reasonnings (if $t_k < t'_k$ and if $t'_k < t_k$),
and Equation \ref{equ:1}, we derive:
\begin{equation} \label{equ:2}
  %\onehalfspacing
  \tallyIdx{i'}{X'}(t') > \tallyIdx{i}{X}(t)
\end{equation}
\noindent
where $\tallyIdx{k}{Z}(\tau)$ means the value of \tally{Z} computed at time
$\tau$ on \SiteId{k}.

Now consider some $k \in Q'$.

If $t_k > {t'}_k$ then
  $X$ being the first candidate non-compatible with
  $X'$ elected in $r$, from Algorithm~\ref{alg:proposal_update},
  we have \prefix{X'}{proposals_i[k](t)}.

If $t_k < {t'}_k$, now either
  \begin{enumerate}
  \item \prefix{X'}{proposals_i[k](t)}
  \item or $k$ has not yet voted on $X.K$ on $i$ at $t$.
  \end{enumerate}
  The reasoning here is similar to $k \in Q$:
  we use the minimality of $X$ and $X'$,
  the fact that they are non-compatible,
  and that $X$ is the first candidate non-compatible with $X'$ elected in $r$.

From the above, it follows that:
\begin{equation}
  \label{equ:3}
  %\onehalfspacing
  \tallyIdx{i'}{X'}(t') < \tallyIdx{i}{X'}(t) + \cotallyIdx{i}{X}(t)
\end{equation}
\noindent
Now, combining equations \ref{equ:2} and \ref{equ:3}, we conclude that,
at site $i$ at time $t$:
\begin{equation}
  \label{equ:4}
  %\onehalfspacing
  \tally{X} < \max\limits_{B \in \opponents{X}}(\tally{B}) + \cotally{X}
\end{equation}

\noindent
$X$ cannot be elected on $i$ at $t$.
Contradiction.

\subsection{Time complexity to run an election }

Let $M$ be a site-multilog, and let $m$ be the number of actions in $M$.
We first extract from $\proposals$ the set of
candidates as follows:
\begin{enumerate}
\item For every proposal $P \in \proposals$, for every actions 
  $\alpha \in P$, we compute the list of predecessors by
  \NotAfter, \Enables and \NonCommuting of $\alpha$ in $P$.
\item Let $l$ be such a list,
  we then compute $d=l \inter \dead{P}$ for every $P$.
\item Then for any couple $(\alpha,\beta) \in l$ such 
  that $\alpha \NonCommuting \beta \in \NonCommuting_{P}$,
  we save the serialization decision:
  either $\NotAfterAB \in \NotAfter_{P}$ or $\NotAfterBA \in \NotAfter_{P}$.
  It forms a set of couples $s$, containing at most $\frac{1}{2}(m^2 - m)$
  elements.%
  \footnote{
    It equals the maximum number of edges in a strongly connected graph of size $m$
  }%
\end{enumerate}
A candidate is any tuple $X=(l,d,s)$.
According to items 1,2 and 3, the time complexity to extract all the candidates in $\proposals$,
is at most $O(nm)$ since all operations can be performed simultaneously.

We compute \cotally{X} by comparing $l$ to $P.K$
for any $P \in \proposals$: $O(mn)$ operations.
Finally we divide the remaning proposals
into \tally{X} and \opponents{X} by comparing
\dead{P} and $\NonCommuting_{P}$ to
$d$ and $s$: $O(n(m+s))$ operations.

Since \NonCommuting is symetric, 
it can exist at most $O(n(m-\sqrt{2s}))$ candidates.
Thus we have to consider the maximum of the function
$(m+s)(m-\sqrt{2s})$.
It follows that $s=\frac{2}{9}m^2$, and that the time complexity
of the whole election process is  $O(m^3n^2)$.

\subsection{Message cost}

Interestingly, the message cost of our protocol varies with application
semantics, along two dimensions.

First, the \emph{degree of semantic complexity}, i.e., the complexity of
the client constraint graph \CM, influences the number of votes
required.
To illustrate, consider an application where all actions are mutually
independent, i.e., \CM contains no constraints.
Then, all actions commute with one another, and no action never needs to
be killed.
Every candidate is trivially eligible, and trivially compatible with all
other candidates.
% **************** SO WHAT??? ****************

Second, call \emph{degree of optimism} $d$ the size of a batch, i.e.,
the number of actions that a site may execute tentatively before
requiring commitment.
This measures both that replicas relax consistency and that clients
propose to the same replica, concurrent commutative actions. % **** ????
It takes a chain of $\frac{n}{2}$ messages to construct a majority.
A candidates may contain up to $d$ actions.
Therefore, the amortised message cost to commit an action is
$\frac{n}{2} \times \frac{1}{d}$.

A more detailed evaluation of message cost
%% and fault resilience
is left for future work.

%% Conversely, an application where all action pairs are non-commuting and
%% there are no antagonisms (hence no actions to be made dead) requires a
%% total order; hence it will have to pay at least the cost of consensus.

%% A related issue is fault tolerance, which is also related to semantics.
%% An application with no constraints whatsoever is trivially fault
%% tolerant, since in this case eventual delivery of site-multilogs is
%% sufficient.

\subsection{Implementation considerations}

Our pseudo-code was written for clarity, not efficiency.
Many optimisations are possible.
For instance, a site $i$ does not need to send the whole \proposalsi[i].
When sending to $j$, it suffices to send the difference $\proposalsi[i]
\setminus \proposalsi[j]$.

Conceptually, a multilog grows without bound.
However, a stable action, and all its constraints, can safely be
deleted.   

Conceptually, our algorithm executes all actions everywhere.
A practical implementation only needs to achieve an equivalent state; in
particular actions that do not have side-effects do not have to be
replayed.
For instance, in a database application, read operations do not to be
replayed.%
\footnote{
  Formally, we need to generalise the equivalence relation between
  schedules, which currently is based only on  \NonCommuting
  \cite{rep:syn:1484}.
  The definition of consistency now becomes that every pair of sites
  eventually converges to schedules that are equivalent according to
  the new relation.
}%

\begin{table}[t]\centering
  \begin{displaymath}
    \begin{array}{c|c|c|c|}
      \multicolumn{1}{c}{~}
                                            &\multicolumn{1}{c}{~~T \HB T'~~}
                                                                       &\multicolumn{1}{c}{~~T \CC T'~~}
                                                                                                &\multicolumn{1}{c}{~~T' \HB T~~} \\
      \cline{2-4} \rule[-1.1ex]{0pt}{4ex}
      \RS{T} \inter  \WS{T'} \neq \emptyset & \NotAfterPair{T}{T'}     & \NotAfterPair{T}{T'} & \CausePair{T'}{T} \\

      \cline{2-4} \rule[-1.1ex]{0pt}{4ex}
      \WS{T} \inter \WS{T'} \neq \emptyset  &  \NotAfterPair{T}{T'} & \NonCommutingPair{T}{T'} & \NotAfterPair{T'}{T}  \\
      \cline{2-4}
    \end{array}
  \end{displaymath}%
  \caption{
    $\CM_{\text{SER-DB-after}}$: Constraints for a serialisable
    database that transmits after-values}
\label{table:its:db}
\end{table}

\subsection{Example application}
\label{sec:application}

We illustrate the application of our algorithm to a replicated database.
The semantic constraints between two transactions depend on several
factors:
% \begin{enumerate}
%  \item
(i)~Whether the transactions are related by happens-before or are
           concurrent.
%  \item
(ii)~Whether their read- and write-sets intersect or not.
%  \item
(iii)~What consistency criterion is being enforced
        (for instance, constraints differ between serializability and
        snapshot isolation \cite{syn:db:1467}).
%  \item
(iv)~How, after executing a transaction on some initial site, the
        system replicates its effects at a
        remote site: by replaying the transaction, or by applying the
        after-values computed at the initial site.  
% \end{enumerate}

Table~\ref{table:its:db} exhibits semantic
constraints between transactions, where (a) the system replicates a
transaction by writing its after-values, and (b) transactions are
strictly serialisable.%
\footnote{
    $T \HB T'$ denotes $T$ happens-before $T$
    \cite{con:rep:615}.
    $T \CC T'$ denotes concurrency, i.e., neither $T \HB T'$, nor $T' \HB
    T$.
    \RS{T} and \WS{T} denote $T$'s read set and write set respectively.
}
Supporting a different semantics, e.g., (a') replaying actions, or (b')
SI, requires only some small changes to the table.

\section{Related work}
\label{sec:related}

In previous OR systems, commitment was often either centralised at a
primary site \cite{syn:1474,syn:rep:1433} or oblivious of
semantics \cite{bd:rep:1454,rep:syn:1500}.
It is very difficult to combine decentralisation with semantics.

Our election algorithm is inspired by Keleher's Deno system
\cite{syn:rep:1440}, a pessimistic system, which performs a discrete
sequence of elections.
Keleher proposes plurality voting to ensure progress when none of
multiple competing proposals gains a majority.
The VVWV protocol of Barreto and Ferreira generalizes Deno's voting
procedure, enabling continuous voting \cite{rep:syn:1565}.

The only semantics supported by Deno or VVWV is to enforce Lamport's
happens-before relation \cite{con:rep:615}; all actions
are assumed be mutually non-commuting.
Happens-before captures potential causality;
however an event may happen-before another even if they are not
truly dependent. This paper further generalizes VVWV by
considering semantic constraints.

Holliday et al.\ depict a family of epidemic algorithms
to ensure serializability in replicated datbase systems \cite{942746}.
The three algorithms consider that
concurrent conflicting transactions are antagonistic.
Two of them abort concurrent conflicting transactions,
and the last one (quorum-based) can only commit one transactions
among a set of concurrent conflicting ones.
Our algorithm consider that concurrent conflicting transactions
are not necessarily antagonistic, it tries to optimize the number
of committed transactions, computing a best-effort proposal ,
and electing them with plurality.

%% In particular, we distinguish causal dependence from happens-before.
%% Using the separate constraints \NotAfter and \Enables we can encode more
%% relations.
%% We allow non-antagonistic concurrent actions to co-exist in a schedule.

%%%Lamport's happens-before relation \cite{con:rep:615} connects events in
%%%a distributed system.

ESDS \cite{alg:rep:syn:1464} is a decentralised replication protocol
that supports some semantics.
It allows users to create an arbitrary
causal dependence graph between actions.
ESDS eventually computes a global total order among actions, but also
includes an optimisation for the case where some action pairs commute.
ESDS does not consider atomicity or antagonism relations, nor does it
consider dead actions.

Bayou \cite{syn:rep:1433} supports arbitrary application
semantics. User-supplied code controls whether an action is
committed or aborted. However the system imposes an arbitrary
total execution order. Bayou centralises decision at a single
primary replica.

IceCube \cite{syn:rep:1432} introduced the idea of reifying semantics
with constraints.
The IceCube algorithm computes optimal proposals, minimizing the number
of dead actions.
Like Bayou, commitment in IceCube is centralised at a primary.
Compared to this article, IceCube supports a richer constraint
vocabulary, which is useful for applications, but harder to reason about
formally.

The Paxos distributed protocol \cite{syn:rep:1548}
computes a total order.
Such total order may be used to implement
\emph{state-machine replication} \cite{con:rep:615}, whereby all
sites execute exactly the same schedule.
Such a total order over
all actions is necessary only if all actions are mutually
non-commuting.
In Section~\ref{sec:classical} we showed how to combine semantic
constraints with a total order, but this approach is clearly
sub-optimal.
Howover, Paxos remains live even if $f < \frac{n}{2}$ sites crash
forever, whereas the other systems described here (including ours) block
if a site crashes forever.
We assume that a site stores its multilogs and its proposals in
persistent memory, and that after a crash it with its identity and
persistent store intact.
This is a fairly reasonable assumption in a well-managed cooperative
system.
(For instance, each site might actually be implemented as a cluster on a
LAN, with redundant storage, and strong consistency internally.)

Generalized Paxos \cite{rep:syn:1567} and Generic Broadcast
\cite{rep:syn:1569}  take commutativity relations  into account
and compute a partial order. They do not consider any other
semantic relations. Both Generalized Paxos \cite{rep:syn:1567} and
our algorithm make progress when a majority is not reached,
although through different means. Generalized Paxos starts a new
election instance, whereas our algorithm waits for a plurality
decision.

\section{Conclusion and future work}
\label{sec:conclusion}

The focus of our study is cooperative applications with rich semantics.
Previous approaches to replication did not support a sufficiently rich
repertoire of semantics, or relied on a centralized point of commitment.
They often impose a total order, which is stronger than necessary.

In contrast, we propose a decentralized commitment protocol for
se\-man\-tic\-ally-rich systems.
Our approach is to reify semantic relations as constraints, which
restrict the scheduling behavior of the system.
According to our formal definition of consistency, the system has an
obligation to resolve conflicts, and to eventually execute equivalent
stable schedules at all sites.

Our protocol is safe in the absence of Byzantine faults, and live in the
absence of crashes.
It uses voting to avoid any centralization bottleneck, and to ensure
that the result is similar to local proposals.
It uses plurality voting to make progress even when an election does not
reach a majority.

There is an interesting trade-off in the proposal{\slash}voting
procedure.
The system might decide frequently, in small increments, so that users
quickly know whether their tentative actions are accepted or rejected.
However this might be non-optimal as it may cut off interesting future
behaviors.
Or it may base its decisions on a large batch of tentative actions,
deciding less frequently.
This imposes more uncertainty on users, but decisions may be closer to
the optimum.
We plan to study this trade-off in our future work.

Another future direction is partial replication.
In such a system, a site receives only the actions relative to the
objects it replicates (and their constraints).
A site votes only on the actions it knows.
Because constraints might relate
actions known only by distinct sites, these sites must agree together;
however we expect that global agreement is rarely necessary.
By exploiting knowledge of semantic constraints, we hope to limit the
scope of a commitment protocol to small-scale agreements, instead of a
global consensus.

    {\small
      \renewcommand{\url}[1]{}
      %\singlespacing
      \bibliographystyle{plain}
      \bibliography{bib,main}

\begin{thebibliography}{10}

\bibitem{rep:syn:1565}
Jo{{\~a}}o Barreto and Paulo Ferreira.
\newblock An efficient and fault-tolerant update commitment protocol for weakly
  connected replicas.
\newblock In {\em Euro-Par}, pages 1059--1068, Lisbon, Portugal, September
  2005.
\newblock \url{http://dx.doi.org/10.1007/11549468_116}.

\bibitem{syn:db:1467}
Philip~A. Bernstein, Vassos Hadzilacos, and Nathan Goodman.
\newblock {\em Concurrency Control and Recovery in Database Systems}.
\newblock Addison-Wesley, 1987.
\newblock \url{http://research.microsoft.com/pubs/ccontrol/}.

\bibitem{alg:rep:1459}
Alan~J. Demers, Daniel~H. Greene, Carl Hauser, Wes Irish, and John Larson.
\newblock Epidemic algorithms for replicated database maintenance.
\newblock In {\em Symp.\ on Principles of Dist.\ Comp.\ (PODC)}, pages 1--12,
  Vancouver, BC, Canada, August 1987.
\newblock Also appears Op.\ Sys.\ Review 22(1): 8-32 (1988).

\bibitem{alg:rep:syn:1464}
Alan Fekete, David Gupta, Victor Luchangco, Nancy Lynch, and Alex Shvartsman.
\newblock Eventually-serializable data services.
\newblock {\em Theoretical Computer Science}, 220(Special issue on Distributed
  Algorithms):113--156, 1999.

\bibitem{942746}
JoAnne Holliday, Robert Steinke, Divyakant Agrawal, and Amr~El Abbadi.
\newblock Epidemic algorithms for replicated databases.
\newblock {\em IEEE Transactions on Knowledge and Data Engineering},
  15(5):1218--1238, 2003.

\bibitem{alg:app:1584}
Claudia-Lavinia Ignat and Moira~C. Norrie.
\newblock {D}raw-{T}ogether: Graphical editor for collaborative drawing.
\newblock In {\em Int.\ Conf.\ on Computer-Supported Cooperative Work (CSCW)},
  pages 269--278, Banff, Alberta, Canada, November 2006.

\bibitem{bd:rep:1454}
Paul~R. Johnson and Robert~H. Thomas.
\newblock The maintenance of duplicate databases.
\newblock Internet Request for Comments RFC 677, Information Sciences
  Institute, January 1976.
\newblock \url{http://www.rfc-editor.org/rfc.html}.

\bibitem{syn:rep:1440}
Peter~J. Keleher.
\newblock Decentralized replicated-object protocols.
\newblock In {\em Symp.\ on Principles of Dist.\ Comp.\ (PODC)}, pages
  143--151, Atlanta, GA, USA, May 1999. ACM Press.
\newblock \url{http://doi.acm.org/10.1145/301308.301345}.

\bibitem{syn:rep:1432}
Anne-Marie Kermarrec, Antony Rowstron, Marc Shapiro, and Peter Druschel.
\newblock The {I}ce{C}ube approach to the reconciliation of divergent replicas.
\newblock In {\em Symp.\ on Principles of Dist.\ Comp.\ (PODC)}, Newport, RI,
  USA, August 2001. ACM SIGACT-SIGOPS, ACM Press.
\newblock
  \url{http://research.microsoft.com/research/camdis/Publis/podc2001.pdf}.

\bibitem{con:rep:615}
Leslie Lamport.
\newblock Time, clocks, and the ordering of events in a distributed system.
\newblock {\em Communications of the ACM}, 21(7):558--565, July 1978.

\bibitem{syn:rep:1548}
Leslie Lamport.
\newblock The part-time parliament.
\newblock {\em ACM Transactions on Computer Systems}, 16(2):133--169, May 1998.
\newblock \url{http://doi.acm.org/10.1145/279227.279229}.

\bibitem{rep:syn:1567}
Leslie Lamport.
\newblock Generalized consensus and {P}axos.
\newblock Technical Report MSR-TR-2005-33, Microsoft Research, March 2005.
\newblock \url{ftp://ftp.research.microsoft.com/pub/tr/TR-2005-33.pdf}.

\bibitem{db:rep:1586}
F.~Pedone, R.~Guerraoui, and A.~Schiper.
\newblock The database state machine approach.
\newblock {\em J.\ of Dist.\ and Parallel Databases and Technology},
  14(1):71--98, 2003.

\bibitem{rep:syn:1569}
Fernando Pedone and Andr{{\'e}} Schiper.
\newblock Handling message semantics with generic broadcast protocols.
\newblock {\em Distributed Computing Journal}, 15(2):97--107, 2002.
\newblock \url{http://www.inf.unisi.ch/faculty/pedone/papers/2002DC.pdf}.

\bibitem{syn:1474}
Nuno Pregui{\c{c}}a, Marc Shapiro, and Caroline Matheson.
\newblock Semantics-based reconciliation for collaborative and mobile
  environments.
\newblock In {\em Proc.\ Tenth Int.\ Conf.\ on Coop.\ Info.\ Sys.\ (CoopIS)},
  volume 2888 of {\em Lecture Notes in Comp.\ Sc.}, pages 38--55, Catania,
  Sicily, Italy, November 2003. {S}pringer-{V}erlag.
\newblock \url{http://www-sor.inria.fr/~shapiro/papers/coopis-2003.pdf}.

\bibitem{sys:rep:1585}
David Ratner, Peter Reiher, and Gerald Popek.
\newblock Roam: A scalable replication system for mobile computing.
\newblock In {\em Int.\ W.\ on Database \& Expert Systems Apps.\ (DEXA)}, pages
  96--104, Los Alamitos, CA, USA, 1999. IEEE Comp.\ Society.
\newblock \url{http://doi.ieeecomputersociety.org/10.1109/DEXA.1999.795151}.

\bibitem{rep:syn:1500}
Yasushi Saito and Marc Shapiro.
\newblock Optimistic replication.
\newblock {\em Computing Surveys}, 37(1):42--81, March 2005.
\newblock \url{http://doi.acm.org/10.1145/1057977.1057980}.

\bibitem{rep:syn:1484}
Marc Shapiro and Karthik Bhargavan.
\newblock The {A}ctions-{C}onstraints approach to replication: Definitions and
  proofs.
\newblock Technical Report MSR-TR-2004-14, Microsoft Research, March 2004.
\newblock \url{ftp://ftp.research.microsoft.com/pub/tr/TR-2004-14.pdf}.

\bibitem{alg:rep:1460}
Chengzheng Sun, Xiaohua Jia, Yanchun Zhang, Yun Yang, and David Chen.
\newblock Achieving convergence, causality preservation, and intention
  preservation in real-time cooperative editing systems.
\newblock {\em Trans.\ on Comp.-Human Interaction}, 5(1):63--108, March 1998.
\newblock \url{http://doi.acm.org/10.1145/274444.274447}.

\bibitem{syn:rep:1433}
Douglas~B. Terry, Marvin~M. Theimer, Karin Petersen, Alan~J. Demers, Mike~J.
  Spreitzer, and Carl~H. Hauser.
\newblock Managing update conflicts in {B}ayou, a weakly connected replicated
  storage system.
\newblock In {\em 15th Symp.\ on Op.\ Sys.\ Principles (SOSP)}, pages 172--182,
  Copper Mountain, CO, USA, December 1995. ACM SIGOPS, ACM Press.
\newblock
  \url{http://www.acm.org/pubs/articles/proceedings/ops/224056/p172-terry/p172%
-terry.pdf}.

\end{thebibliography}
    }

\end{document}